\documentclass[amssymb,aps,prb,floats,twocolumn,superscriptaddress,showpacs]{
revtex4-1}
\usepackage{graphicx}
\usepackage{graphics}
\usepackage{bm}
\usepackage{tabularx}
\usepackage{array}
\usepackage{float}
\usepackage[T1]{fontenc}

\graphicspath{{images/ultimas/}}

\begin{document}

\title{Large perpendicular magnetic anisotropy in 
magnetostrictive Fe$_{1-x}$Ga$_x$ thin films}

\author{M. Barturen}
\affiliation{CNEA-CONICET, Centro At\'omico Bariloche, (R8402AGP) San Carlos de Bariloche, Argentina}
\affiliation{Instituto Balseiro, Universidad Nacional de Cuyo, Centro At\'omico Bariloche, (R8402AGP) San Carlos de Bariloche, Argentina}
\affiliation{Sorbonne Universit\'es, UPMC Univ Paris 06, UMR 7588, INSP, 4 place Jussieu, F-75005, Paris, France}
\affiliation{CNRS, UMR 7588, Institut des Nanosciences de Paris, 4 place Jussieu, F-75005, Paris, France}
\affiliation{LIFAN, Laboratorio Internacional Franco-Argentino en Nanociencias}

\author{J. Milano}
\affiliation{CNEA-CONICET, Centro At\'omico Bariloche, (R8402AGP) San Carlos de Bariloche, Argentina}
\affiliation{Instituto Balseiro, Universidad Nacional de Cuyo, Centro At\'omico Bariloche, (R8402AGP) San Carlos de Bariloche, Argentina}
\affiliation{LIFAN, Laboratorio Internacional Franco-Argentino en Nanociencias}

\author{M. V\'asquez-Mansilla}
\affiliation{CNEA-CONICET, Centro At\'omico Bariloche, (R8402AGP) San Carlos de Bariloche, Argentina}

\author{C. Helman}
\affiliation{LIFAN, Laboratorio Internacional Franco-Argentino en Nanociencias}
\affiliation{CNEA, Centro At\'omico Constituyentes, Avenida General Paz 1499, 
San Mart\'{\i}n, Argentina}

\author{M. A. Barral}
\affiliation{LIFAN, Laboratorio Internacional Franco-Argentino en Nanociencias}
\affiliation{CNEA, Centro At\'omico Constituyentes, Avenida General Paz 1499, 
San Mart\'{\i}n, Argentina} 

\author{A. M. Llois}
\affiliation{LIFAN, Laboratorio Internacional Franco-Argentino en Nanociencias}
\affiliation{CNEA, Centro At\'omico Constituyentes, Avenida General Paz 1499, 
San Mart\'{\i}n, Argentina}

\author{M. Eddrief}
\affiliation{Sorbonne Universit\'es, UPMC Univ Paris 06, UMR 7588, INSP, 4 place Jussieu, F-75005, Paris, France}
\affiliation{CNRS, UMR 7588, Institut des Nanosciences de Paris, 4 place Jussieu, F-75005, Paris, France}
\affiliation{LIFAN, Laboratorio Internacional Franco-Argentino en Nanociencias}

\author{M. Marangolo}
\affiliation{Sorbonne Universit\'es, UPMC Univ Paris 06, UMR 7588, INSP, 4 place Jussieu, F-75005, Paris, France}
\affiliation{CNRS, UMR 7588, Institut des Nanosciences de Paris, 4 place Jussieu, F-75005, Paris, France}
\affiliation{LIFAN, Laboratorio Internacional Franco-Argentino en Nanociencias}

\date{\today}

\begin{abstract} In this work we report the appearence of a large perpendicular 
magnetic anisotropy (PMA) in Fe$_{1-x}$Ga$_x$ thin films grown onto 
ZnSe/GaAs(001). This arising anisotropy is related to the tetragonal 
metastable phase in as-grown samples recently reported [M. Eddrief 
{\it et al.}, Phys. Rev. B {\bf 84}, 161410 (2011)]. By means of ferromagnetic 
resonance studies we measured PMA values up to $\sim$~5$\times$10$^5$~J/m$^3$. 
PMA vanishes when the cubic structure is recovered upon annealing at 
300$^{\circ}$C. Despite the important values of the magnetoelastic constants 
measured via the cantilever method, the consequent magnetoelastic contribution 
to PMA is not enough to explain the observed anisotropy values in the distorted 
state. {\it Ab initio} calculations show that the chemical ordering plays a 
crucial role in the appearance of PMA. Through a phenomenological model we 
propose that an excess of next nearest neighbour Ga pairs (B$_2$-like 
ordering) along the perpendicular direction arises as a possible source of PMA 
in Fe$_{1-x}$Ga$_x$ thin films.
\end{abstract}
\maketitle

\section{Introduction}

Nowadays, active research efforts are being carried out to stabilize 
perpendicular magnetic anisotropy (PMA) in thin films and nanostructures 
because it is at the intersection of different research 
streams of modern magnetism and spintronics. An example is the spin 
transfer torque - magnetic random access memory (STT-MRAM) where magnetized 
units are more easily switched by a lower electrical current when the 
electrodes present PMA.\cite{Mangin2006,yakata}\\
The PMA tends to align the magnetic moments within the film perpendicular to 
the film surface. To achieve this goal, different approaches have been adopted 
based on very different physical phenomena.
PMA can be obtained by magnetocrystalline anisotropy (MCA) (such as in 
Co),\cite{cullity} interface effects,\cite{Ikeda2010} stress\cite{Butera2010,Schulz1994} 
and shape of the magnetic entities.\cite{Vidal2012} In particular, 
magnetocrystalline effects can be used for obtaining 
PMA in materials with a preferential crystalline axis. This is the case of Co 
when its easy axis is along the film growth direction (perpendicular to the 
film surface).\cite{PhysRevB.stripesCo} Furthermore, it is possible to 
manipulate the crystallographic structure and to 
induce an easy axis of the magnetization by adjusting the growing conditions or 
postdeposition treatments. This is the case of the L1$_{0}$-ordered alloys, 
whose particular chemical ordering is reached by thermal annealing or 
by high temperature thin film growth.\cite{chang} The tetragonal distortion 
induced by the atomic arrangement is responsible for the 
appearence of the PMA.\cite{devdatstor} An example of these alloys is FePt 
which presents a very high PMA, 
i.e. $\sim$3-7$\times$10$^6$~J/m$^3$.\cite{Ivanov1973,Seki2006} However, it is 
worthwhile to report that more "moderate" PMA values are needed to obtain lower 
switching fields in recording applications. This motivates the study of the 
L1$_{0}$-ordered FePd alloys where PMA 
is $\sim$1$\times$10$^6$~J/m$^3$.\cite{weller}  

In this work, we report the presence of PMA in 
Fe$_{1-x}$Ga$_x$/ZnSe/GaAs(001) thin films. A combined experimental and 
theoretical study of such films as a 
function of the Ga content and the atomic ordering was performed. The PMA was 
deduced from ferromagnetic resonance (FMR) experiments and large values up to 
$\sim$~5$\times$10$^5$~J/m$^3$ are found. Due to demagnetization fields, the 
magnetization lies in the film plane of our samples; however the measured PMA 
values assure that the magnetization is perpendicular to the film plane 
when the lateral size is smaller than $\sim$130~nm.\cite{dcr} 
By taking inspiration from a phenomenological model proposed for 
bulk Fe$_{1-x}$Ga$_x$,\cite{Cullen-Ga-Ga-pairs} we estimate the contribution 
of Ga ordering to magnetic anisotropy energy (MAE) due to PMA. This permits us 
to suggest that the strong PMA is due to a preferential 
Ga ordering along the growth direction. Our first-principles calculations in 
the framework of density functional theory account for the role played by Ga 
ordering.  
The experimental observation of 
"moderate" PMA is of considerable interest for the following reasons: (i) 
Fe$_{1-x}$Ga$_x$ (Galfenol) is a material well known for its very high 
magnetostrictive coefficient and its growth is well mastered by today's 
industry; 
(ii) Epitaxial films of Fe$_{1-x}$Ga$_x$ on GaAs present spintronic properties 
like spin injection into a semiconductor,\cite{vanterve} strain-tuned 
magnetic anisotropy and high frequency properties in the 
microwave region;\cite{Parkes} (iii) 
PMA is obtained without any annealing procedure and at moderate growth 
temperature compatible with ferromagnetic/semiconductor heterostructures 
engineering. 
In addition to all of these avantages, at present, Ga has a lower cost than Pt 
or Pd due to its crustal abundance.\cite{coeyIEEE} 

\begin{figure}[ht]
 \begin{center}
\includegraphics[width=1\linewidth]{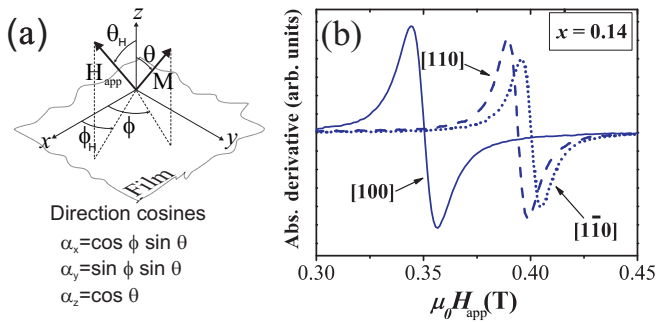}
  \caption{(Color online) (a) Coordinate system used in Eqs.~(\ref{eq:U}) and 
(\ref{eq:Smit-Beljers}). (b) FMR spectra for different directions of $H_{\rm 
app}$ for the Fe$_{0.86}$Ga$_{0.14}$ sample.}
  \label{fig1:comp}
 \end{center}
\end{figure}

\section{Experimental}

Epitaxial Fe$_{1-x}$Ga$_x$ samples were grown by Molecular Beam Epitaxy on 
c(2x2) Zn-terminated ZnSe epilayers onto GaAs(001) 
substrates.\cite{PhysRevLett.Max,eddriefPRB2006} At the end, the films were 
covered by a 
protective 3-nm gold capping layer. Details of the growth are given in 
Ref.~\onlinecite{ref:Eddrief-PRB}. We fabricated 72-nm nominal thick samples at 
several 
Ga concentrations, $x$~=~0.14, 0.18, 0.20. Such concentrations were determined 
by means of x-ray photoelectron spectroscopy (XPS) and confirmed by Rutherford 
backscattering (RBS) and energy dispersive x-ray spectrometry (EDX). 
With the aim to study the 
response to thermal treatment, a selection of samples of varying Ga 
concentration (for $x$~=~0.14 and 0.20) were 
annealed at 300$^{\circ}$C in ultra high vacuum\cite{ref:Eddrief-PRB} as well. 
This annealing temperature is sufficient to cause Ga mobility in the Fe matrix, 
as attested in Ref.~\onlinecite{kubaschewski} and, at the same time, it 
preserves the 
sharp FM/SC interface.\cite{PhysRevLett.Max}
By X-ray diffraction (XRD), we determine the lattice parameters and we 
observe a Ga dependent tetragonal deformation. Ref.~\onlinecite{ref:Eddrief-PRB} 
suggests that this deformation is due to Ga pairing 
along [001] direction leading to a local B$_2$-like structure as forseen by 
previous {\it ab initio} studies\cite{Zhang20114044,WuJAP2002} and 
experimentally observed by differential x-ray absorption spectroscopy 
(DiffXAS).\cite{Ruffoni2008} Such a B$_2$-like phase occurs when Ga pairs are 
formed at next nearest neighbours along the <100> family direction. The first 
magnetic characterizations put in evidence clearly the presence of PMA where 
the formation of stripe-like domains is 
reported.\cite{BarturenAPL,BarturenEPJB} We performed 
FMR experiments in a 
Bruker ESP-300 spectrometer at $\nu$~$\sim$~24~GHz (K-band). The angular 
dependence of the resonance spectra has been studied in the in-plane geometry 
by fixing the polar angle at $\theta_H$~=~90$^{\circ}$ while varying $\phi_H$ 
from 0$^{\circ}$ to 360$^{\circ}$ [see Fig.~\ref{fig1:comp}(a)]. We also measured 
the magnetoelastic coupling by performing cantilever deflection experiments. The 
magnetic field was applied along the [100] direction in order to obtain the 
magnetoelstic coupling along such direction. Additional magnetization 
measurements have been performed in a superconducting quantum interference 
device (SQUID) and a vibrating sample magnetometer (VSM).

\section{Results and Discussion}

In order to study quantitatively the magnetic anisotropies in our samples 
through FMR, we evaluate how these anisotropies contribute to the resonance 
field, $H_r$. To do this, we propose a self-consistent scheme 
that solves the equilibrium position of $\bf{M}$ via the magnetic free energy 
$U$ and a linearized version of the Landau-Ginzburg equation of motion for the 
magnetization,\cite{Smit-Beljers} when the sample is magnetically saturated. 
It is important to notice that the onset of a tetragonal distortion along 
$z$-axis modifies the well-known cubic anisotropy energy term of pure Fe thin 
films, i.e. the cubic term [$K_4^{\rm 
Fe}(\alpha_x^2 \alpha_y^2+\alpha_y^2 \alpha_z^2+\alpha_z^2 \alpha_x^2)$] is 
modified as $K_4^{\rm IP} \alpha_x^2 \alpha_y^2+K^{\rm OP} \alpha_z^2$, where 
the out-of-plane $\alpha_z^4$-term was neglected.
$\alpha_i$ denotes the direction cosines of the magnetization 
[Fig.~\ref{fig1:comp}(a)] and IP (OP) is the
in-plane (out-of-plane) direction. Hence, the proposed 
expression for $U$ in our coordinate system 
[Fig.~\ref{fig1:comp}(a)] is the following:
\begin{eqnarray}\label{eq:U}
 U(\theta,\phi) = &- & \mu _0\textbf{H}_{\rm app} \cdot \textbf{M} + 
\frac{\mu_0}{2} M^2 \cos^2 \theta \\ \nonumber
& + & \frac{1}{4}K^{\rm IP}_{4} \sin ^4\theta \sin^2 2\phi - K_{\rm PMA} \cos 
^2 \theta \\ \nonumber
& + & K_u \sin^2\theta 
\cos^2 \left( \phi - \frac{\pi}{4}\right)
\end{eqnarray}
\noindent where $\textbf{H}_{\rm app}$ is the applied magnetic field, 
$\textbf{M}$ is the magnetization vector, $\mu _0$ is the vacuum permeability 
($\mu _0$~=~4$\pi\times$10$^{-7}$ T\,m/A). The first term on the right-hand 
side is 
the classical Zeeman energy. The second one is the energy related to the 
demagnetizing dipolar field. The third term is the first order in-plane 
contribution to the magnetocrystalline energy for a 
tetragonal lattice. The fourth one is the energy related to the PMA, 
including $K^{\rm OP}$ and other contributions that will be discused below. 
Finally, the fifth term stands for an uniaxial in-plane anisotropy arising 
from the Fe$_{1-x}$Ga$_x$/ZnSe interface. On the other hand, 
the equation that accounts for the magnetization dynamics in the small 
oscillation approximation was given by Smit and Beljers\cite{Smit-Beljers} and 
it writes: 
\begin{equation}\label{eq:Smit-Beljers}
\left . \omega^2= \frac{\gamma^2}{M^2 \textnormal{sin}^2 
\theta}\left[\frac{\partial^2 
U}{\partial^2 \theta }\frac{\partial^2U}{\partial^2 \phi 
}-\left(\frac{\partial^2 U}{\partial \theta \partial \phi}\right)^2\right] 
\right \arrowvert_{\theta_{eq},\phi_{eq}}, 
\end{equation}
\noindent evaluated at the equilibrium angles, $\theta_{eq}$ and $\phi_{eq}$ 
obtained from Eq.~(\ref{eq:U}), where $\gamma$~$=g\mu_B/\hbar$ is the 
gyromagnetic ratio and $\mu_B$ is the Bohr magneton. The $g$ value was set to 
2.1, as for Fe. 

\begin{figure}[ht]
 \begin{center}
 \includegraphics[width=1\linewidth]{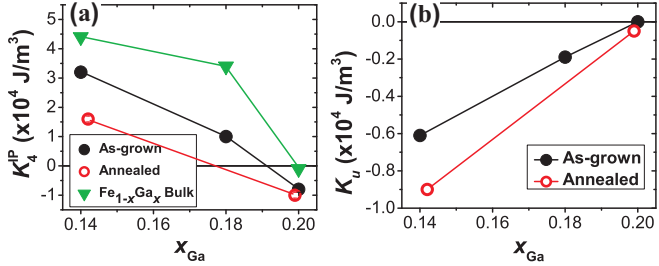} 
  \caption{(Color online) (a) $K_4^{\rm IP}$ and (b) $K_u$ as a function of 
$x$.}
  \label{figKip}
 \end{center}
\end{figure}

In Fig.~\ref{fig1:comp}(b), we show an example of spectra taken at different 
$H_{\rm app}$ in-plane directions. The shift of the resonance mode is due to 
the magnetic anisotropies present in the samples. It is worth to note that 
even though $H_{\rm app}$ is applied only in the film plane, it is possible 
to extract information of the perpendicular anisotropy.\cite{expFMR} The 
magnetic parameters, i.e. $K_{\rm PMA}$, $K^{\rm IP}_4$ and $K_u$, are 
determined by solving self-consistently the Eqs.~(\ref{eq:U}) and 
(\ref{eq:Smit-Beljers}) as mentioned above.
In Fig.~\ref{figKip}(a), we show $K^{\rm IP}_4$ as a function of Ga 
concentration. The $K^{\rm IP}_4$ behavior is similar to 
the one observed in Fe$_{1-x}$Ga$_x$ bulk:\cite{Cullen-Ga-Ga-pairs} (i) It 
decreases for increasing Ga concentrations and (ii) it changes its 
signs.
In Fig.~\ref{figKip}(b), $K_u$ vs. $x$ is shown. This 
contribution arises from the dangling bonds at the 
Fe$_{1-x}$Ga$_x$/ZnSe interface as already seen in Fe/ZnSe 
interfaces.\cite{Sjostedt2002, Marangolo2004} $K_u$ 
diminishes for increasing $x$. This behavior is probably due 
to the fact that amount of Fe atoms at the interface decreases for higher 
Ga concentrations.

\begin{figure}[ht]
 \begin{center}
 \includegraphics[width=1\linewidth]{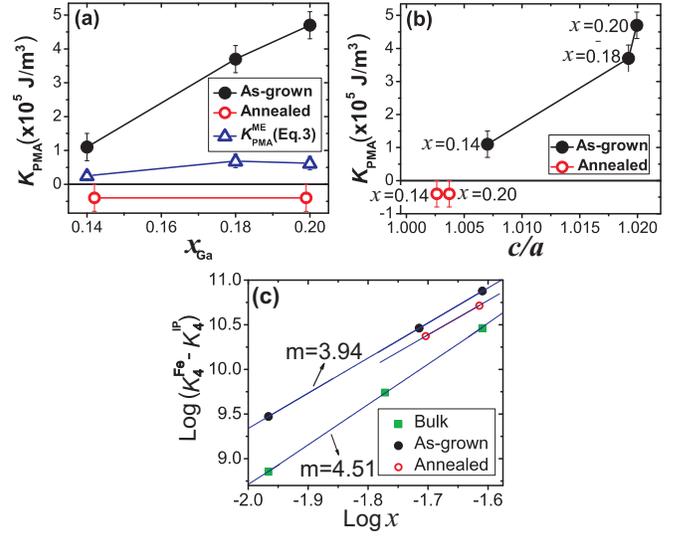} 
  \caption{(Color online) (a) $K_{\rm PMA}$ as a function of $x$. (b)$K_{\rm 
PMA}$ as a 
function of $c/a$. (c) Log-Log plot of ($K^{Fe}_4$~-~$K^{\rm IP}_4$) $vs.$ $x$, 
where the slopes show explicitly the $x^4$ dependence. The $K_4^{\rm IP}$ 
for the cubic symmetry dependence of bulk Fe$_{1-x}$Ga$_x$ taken from 
Ref.~\onlinecite{Cullen-Ga-Ga-pairs} is shown as well.}
  \label{figanipma}
 \end{center}
\end{figure}

In the following, we focus our analysis on 
$K_{\rm PMA}$ and Fig.~\ref{figanipma}(a) displays the best fitted 
values for it.
It is important to note the difference between the results obtained for the 
as-grown and the annealed samples. On one hand, the as-grown samples show very 
high values of $K_{\rm PMA}$ being the highest value 
$\sim$5$\times$10$^5$~J/m$^3$ at $x$~=~0.20, i.e. 10 times larger than pristine 
Fe thin films. On the other hand, for the annealed samples, $K_{\rm PMA}$ shows 
values close to zero. This indicates that the appearence 
of $K_{\rm PMA}$ is related to the tetragonal deformation. 
Fig.~\ref{figanipma}(b) displays $K_{\rm PMA}$ as a function of the $c$/$a$ 
ratio (which measures the tetragonal deformation), where the correlation 
between them is explicitly shown.

In what follows, we discuss possible sources of PMA for our systems. In 
distorted ferromagnetic 
systems, a contribution that cannot be discarded is the one arising from 
magnetoelasticity, specially if the studied system is highly magnetostrictive 
as 
in the case of Fe$_{1-x}$Ga$_x$. Hence, the magnetoelastic contribution to 
$K_{\rm PMA}$ of an out-of-plane tetragonally distorted 
($\epsilon_{yy}$=$\epsilon_{xx}$) film writes:\cite{kittelRevModPhys}
\begin{equation}
K^{\rm ME}_{\rm PMA}=B_1(\epsilon_{zz}-\epsilon_{xx}),
\label{Kpma}
\end{equation}
where $B_1$ is the magnetoelastic constant related to tetragonal strains. 
$\epsilon_{zz}$ ($\epsilon_{xx}$) is the strain along the $z$ ($x$) direction. 
In order to determine if $K_{\rm PMA}$ in our distorted systems 
comes from the magnetoelastic coupling, we evaluate Eq.~(\ref{Kpma}) for a 
given concentration. We consider $\epsilon_{zz}$ ($\epsilon_{xx}$) being equal 
to 
$\frac{c-a_{\rm cub}}{a_{\rm cub}}$ ($\frac{a-a_{\rm cub}}{a_{\rm cub}}$), where 
$a_{\rm cub}$ is the cubic 
lattice parameter extracted from the 
stabilized cubic phase (annealed samples).\cite{expepsilon} $B_1$ was measured 
at MPI (Halle, Germany) for the cubic samples by performing cantilever 
deflection experiments,\cite{Premper_Sander_2012,ref:rev-Sander} which provide 
a direct measurement of the magnetoelastic coupling coefficients in our films. 
The maximum 
value reported by us is $B_1$~=~(5$\pm$1)$\times$10$^6$~J/m$^3$ for $x$~=~0.14. 
In Fig.~\ref{figanipma}(a), we show the calculated $K^{\rm ME}_{\rm PMA}$ via 
Eq.~(\ref{Kpma}). We observe that $K_{\rm PMA}^{\rm ME}$ is only a small 
fraction of $K_{\rm PMA}$ and, therefore, the PMA anisotropy observed by FMR 
does not arise from a simple structural deformation. Consequently, we conclude 
that another source of PMA dominates in our system.
In that sense, we propose that the observed PMA has a magnetocrystalline origin 
linked to the preferential Ga ordering along the [001] direction. Indeed, it is 
well known that Ga-ordering strongly affects MCA in bulk Fe$_{1-x}$Ga$_x$. 
Cullen {\it et al.} have shown the Ga-pairing along <100> directions (second 
nearest neighbors) lowers the magnetocrystalline anisotropy of pristine bulk 
iron ($K_4^{\rm Fe}$) in the following way:
\begin{equation}
\label{Kx4}
K^{\rm IP}_4 = K^{\rm Fe}_4~-~A K^2 x^4, \; {\rm with} \;A~=~4a^2/\pi A_0. 
\end{equation}
Where $K_4$ is the magnecrystalline anisotropy observed in 
Fe$_{1-x}$Ga$_x$, $A_0$ is the exchange stiffness constant and $a$ is related 
to the size of the Ga arrangement, in our case equals one lattice parameter. 
Such $a$ parameter ``measures'' how the Ga atoms affect the magnetic properties 
of the surrounding Fe atoms.
$K$ gives account for a local uniaxial anisotropy due to
the presence of the Ga pairs. Here, we take 
$A_0$~=~1.6$\times$10$^{-11}$~J/m$^3$ consistently 
with reported values for similar films.\cite{Tacchi_rotatable_2014} The 
core of this phenomenological model is that 
such Ga pairs aligned along the <100> axes lead to an extra contribution to the 
free energy that takes into account the tendency for the spin of the neighboring 
Fe atoms to align, in principle, parallel or perpendicular to the Ga-Ga pair 
axes. This energy term reads as follows:
\begin{equation}
\label{deltaE}
U^0_{\rm pair} = \frac{K}{N} \sum_l \sum_i{\alpha_i^2(R_l) 
P_{0,i}(R_l)}. 
\end{equation}
Where $N$ is the number of sites, $\alpha_i(R_l)$ indicates the local 
$R_l$-site magnetization directions with respect to the $i^{\rm th}$ 
($i$~=~$x$,$y$,$z$) axis and $P_{0,i}(R_l)$ is the probability (0 or 1) that 
the Ga 
pairs in the $R_l$ site are aligned along the $i^{\rm th}$ axis. Eq.~(\ref{Kx4}) 
can be also recovered from Eq.~(\ref{deltaE}) in a mean-field 
approach\cite{Cullen-Ga-Ga-pairs,expK2}. Results obtained in 
Ref.~\onlinecite{Cullen-Ga-Ga-pairs}
for bulk Fe$_{1-x}$Ga$_x$ (which will be useful to compare to our results) are 
shown in Fig.~\ref{figanipma}(c) where a $\sim$~$x^4$ dependence is observed. 
A fitting procedure gives $K$~$\sim$~2.9$\times$10$^7$ J/m$^3$. 

In the following we will show how, given an anisotropic distribution of Ga 
pairs along the three <100> direction, the previously reported $K$ values can 
lead to the observed PMA. 
As a starting point we propose that the Ga pairs are preferentially aligned 
along the [001] out of plane direction rather than in plane. This hypothesis is 
supported by the experimentally observed tetragonal distortion leading to the 
B$_2$-like structure mentioned before, coherently with {\it ab initio} 
calculations.\cite{Zhang20114044,WuJAP2002} Consequently, we propose that the 
total probability of finding a Ga pair at the $R_l$-site can be expressed as 
\begin{equation}
P_i(R_l)=P_{0,i}(R_l)+P_z(R_l) \; \delta_{iz},
\end{equation}
where $P_z$ accounts for the 
additional probability of finding Ga pairs along the $z$ axis, which is 
expressed through $\delta_{iz}$. We notice that for cubic Fe$_{1-x}$Ga$_x$ 
(bulk or annealed samples), $P_z$~=~0. On the other hand, for a tetragonal 
distorted crystal 
Eq.~(\ref{deltaE}) becomes:
\begin{equation}
\label{deltaEm}
U^z_{\rm pair} = U^0_{\rm pair} + \frac{K}{N} \sum_l{\alpha^2_z(R_l) P_z(R_l)}. 
\end{equation}
Under this scheme the PMA anisotropy is obtained as 
\begin{eqnarray}
K_{\rm PMA}&=&U_{\rm pair}(\alpha_z=1,P_{0,i},P_z)-U_{\rm pair} (\alpha_z = 
0,P_{0,i},P_z) \nonumber \\ 
&=&K P_z , \;\; {\rm where}\;\; P_z=\frac{\sum_l P_z(R_l)}{N}. \\ \nonumber
\end{eqnarray}

In Fig.~\ref{figanipma}(c), we display log($K^{\rm Fe}_4$~-~$K^{\rm IP}_4$) 
{\it vs.} log~$x$ where the dependency on the fourth power of the Ga 
concentration corroborates the extention of the Cullen's model to the 
tetragonally distorted thin films. A similar behavior is recovered for the 
two annealed samples, as well. 
The important point is that the in-plane measurements permit us to estimate 
 $K$~$\sim$~3.6$\times$10$^7$ J/m$^3$, slightly higher 
than in bulk, adopting the same $A$ value.

For $x$~=~0.20, a lower limit for $<\!P_{0,i}\!>$ can be calculated using
the results of the {\it ab initio} molecular dynamics simulations
performed in Ref.~\onlinecite{Wang2013SciRep}, from which it is possible to
determine the amount of B$_2$-like pairs for a given concentration.
Those calculations foresee that the number of B$_2$-pairs is very small, i.e. 
only 1\% of the D0$_3$ (the majority phase existing in Fe$_{1-x}$Ga$_x$ alloys) 
pairs. An upper limit is obtained by considering a fully random
distribution of the Ga atoms into the bcc Fe matrix ($x^2$), i.e. 
$<\!P_{0,i}\!>$~$\sim$~$x^2$~$\sim$~0.04.\cite{expK2}  
It is important to notice that X-ray diffraction 
experiments\cite{ref:Eddrief-PRB} attest the presence of superlattice 
reflections (B$_2$ or D0$_3$) but the absence of the DO$_3$-characteristic [113] 
reflections. Consequently we estimate that the aforementioned lower limit is not 
encountered in our metastable thin films and that $P_{0,x}+P_{0,y}+P_{0,z}$ is 
close to the upper limit, i.e. $\sim$~10\%. Moreover, the observed tetragonal 
distortion suggests an anisotropic distribution of Ga atoms, with a preferential 
axis along [001]. Even if we are not able to measure the exact Ga-Ga pairs 
distribution in our films, we notice that the $K$ value estimated above is so 
huge that a small $P_z$ unbalance of few 10$^{-2}$ can lead to a K$_{\rm PMA}$ 
close to the observed experimental values of 5$\times$10$^5$ J/m$^3$, for 
$x$~=~0.20. This phenomenological approach allows us to conclude that $K_{\rm 
PMA}$ can be favored if there is an anisotropic distribution of Ga pairs between 
in-plane and out-of-plane directions. However, from this approximation it is not 
possible to determine which is the preferential orientation of the pairs that 
gives rise to the PMA.

\begin{figure}[H]
 \begin{center}
 \includegraphics[width=\linewidth,keepaspectratio=true]{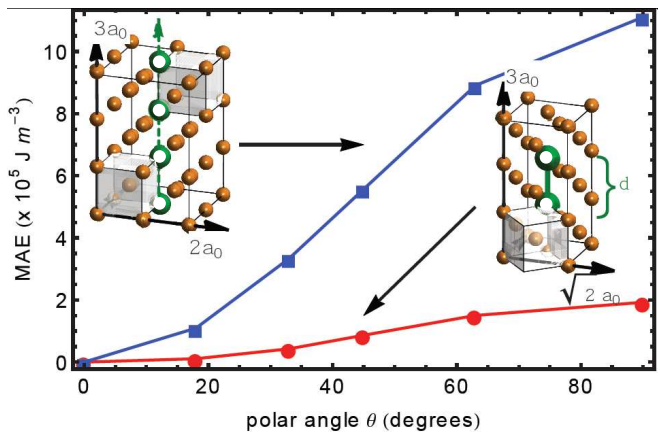}
  \caption{(Color online) MAE as a function of the polar angle, $\theta$, for 
the both calculated structures. Large green (open) and small yellow (filled) 
symbols are Ga and Fe atoms, respectively.}
  \label{figMAE}
  \end{center}
\end{figure}

In this context, from a more microscopical point of view, {\it ab initio} 
calculations within the framework of density functional theory (DFT) can 
provide an insight into the Ga pairing effect on PMA in order to: (i) predict 
the direction of the local easy axis with respect to the Ga pairs direction; 
(ii) estimate the intensity of MAE induced by Ga atoms alignment. The MAE is 
calculated, using a second variational approach as implemented in the Wien2k 
package.\cite{wien2k,expcalc} The pairs are built by Ga atoms located at next 
nearest neighbour positions in the bcc lattice of Fe and directed along [001] 
direction. The in-plane lattice constant, perpendicular to the Ga pairs, is set 
equal to the Fe lattice one, as experimentally observed in our samples 
(see Ref.~\onlinecite{ref:Eddrief-PRB}). The system is allowed to relax in the 
out-of-plane direction.
As we stated above the dependence of $K$ with the 
arrangement of Ga pairs ($a$ in Eq.~\ref{Kx4}) has to be considered. In order 
to take that into account in the calculations, we proposed two situations 
where the Ga pairs are along [001] but are placed in different manners as it is 
shown in the insets of Fig.~\ref{figMAE}. In 
one case single pairs arrangement is assumed (right inset); 
in the second case, the Ga pairs build chains (left inset). The Ga 
concentrations of both structure are similar to the ones measured in the 
studied samples, i.e. $x \sim$~0.16 and $x \sim$~0.12 for the first and the 
second case respectively. The evolution of the MAE as a function of polar 
angle, from the [001] to the [100] directions is shown in Fig.~\ref{figMAE}. 
The MAE for a general orientation of the magnetization is obtained from the 
total energy  difference  with respect to the energy corresponding to the 
magnetization lying along the dimers' direction, i.e. [001]. 
For both Ga distributions treated, the easy axis coincides with the orientation 
of the Ga pairs, that is the [001]. The value of the MAE for the system with 
a chain-like distribution is 5-6 times larger 
than the one obtained for the single pairs B$_2$-like arrangement.  
If we compare the calculated results with ones obtained by us, we observe 
that the measured $K_{\rm PMA}$ values are between the MAE ones 
obtained for both calculated structures.
The big difference between the values of the MAE obtained in the two 
situations analysed shows that the chemical ordering plays a critical role in 
the determination of $K_{\rm PMA}$.
Finally, these {\it ab initio} results represent a strong evidence pointing 
towards the conclusion that the experimentally observed PMA would arise from a 
preferential orientation of the Ga pairs along the $z$-axis, that is, 
perpendicular to the film plane.

\section{Conclusions}

In conclusion, we have demonstrated the presence of a large PMA in 
Fe$_{1-x}$Ga$_x$ thin films grown on ZnSe/GaAs(001), which is correlated to the 
tetragonal distortion reported in such films. The anisotropic distribution of 
Ga pairs along the sample appears as a possible origin of such as PMA. By 
performing {\it ab initio} calculations we show that an excess of such pairs 
along the perpendicular film plane direction provokes the PMA. In order to 
evaluate the Ga pairs contribution to the total magnetic anisotropy, we adopted 
a phenomenological approach similar to the model developed for bulk 
Galfenol,\cite{Cullen-Ga-Ga-pairs} in which an additional energy term, that 
takes into account the preferential axis created by the Ga pairs, was 
introduced. We show that this Ga-pairs term is so important that a slight 
preference of Ga pairs to be aligned along the growth direction could lead to 
the observed $K_{\rm PMA}$. Also, we have studied the in-plane anisotropies. On 
one hand, the magnetocrystalline one presents a similar behavior than 
the one observed in Fe$_{1-x}$Ga$_x$ bulk. On the other hand, the interfacial 
uniaxial anisotropy observed in Fe/ZnSe is also present in our samples. Finally, 
we show that Fe$_{1-x}$Ga$_x$ grown as thin film on GaAs(001) becomes a 
potential material to be used in PMA-based devices, in which magnetostrictive 
and magnetic anisotropy are combined. This could open new perspectives to 
control PMA-related remarkable properties simply by strain.

\begin{acknowledgments}
We thank to Dr. D. Sander and Prof. J. Kirschner for the help in magnetoelastic 
measurements, for valuable discussions and also for the careful reading of this 
manuscript. We acknowledge partial support from CONICET (PIP-00258), ANPCyT 
(PICT-2010-0773) and ANR (SPINSAW, ANR 13-JS04-0001-01).
\end{acknowledgments}


\begin{thebibliography}{43}
\bibitem [{exp({\natexlab{a}})}]{expFMR}%
  \BibitemOpen
  \href@noop {} {} \bibinfo {note} {The resonance conditions
  during a FMR experiment depend on the magnetic anisotropy profile along the
  perpendicular directions respect to applied field at saturation. It 
allows us  to determine the PMA even if the magnetic field is applied in the 
sample plane.}\BibitemShut {Stop}%
\bibitem [{exp({\natexlab{b}})}]{expepsilon}%
  \BibitemOpen
  \href@noop {} {} \bibinfo {note} {In order to determine $\epsilon_{zz}$, we 
  propose that deformation is driven by elastic forces and that the cell 
  volume is preserved as experimentally observed 
  in Ref.~\onlinecite{ref:Eddrief-PRB}}\BibitemShut {NoStop}%
\bibitem [{exp({\natexlab{a}})}]{expK2}%
  \BibitemOpen
  \href@noop {} {} \bibinfo {note} {By simple symmetry considerations, we can 
deduce how $U^0_{\rm pair}$ corrects $K^{\rm Fe}_4$. To do that 
we have to take into account that $K_4$ depends on the fourth power of 
direction cosines, as shown in Eq.~(\ref{eq:U}), while $U_{\rm pair}$ is a 
summatory of terms that depend on $\alpha_i^2$. Then, to respect the $K_4$ four 
fold symmetry given by the direction cosines, the correction from $U^0_{\rm 
pair}$ has to follow a $(<\!P_{0,i}\!> K)^2$ 
dependence. Where $<\!P_{0,i}\!>$ is the average of $P_{0,i}(R_l)$ over the all 
sites and it scales as $x^2$, i.e. the probability to have a neighboring Ga 
atom ($x$) along the $i$-axis (2$x$) time a 1/2 factor in order to avoid double 
counting.}\BibitemShut {Stop}%
\end{thebibliography}
\end{document}